\documentclass[a4paper,twocolumn]{esapub} % European paper
\usepackage{graphics}
\usepackage{amssymb}
\title{The Neupert effect in solar flares and implications \newline for coronal heating}

\author{A. Veronig}
\affil{Institute for Geophysics, Astrophysics and Meteorology,
University of Graz, Universit\"atsplatz 5, A-8010~Graz, Austria, E-mail: asv@igam.uni-graz.at}
\author{B. Vr\v{s}nak}
\affil{Hvar Observatory, Faculty of Geodesy, University of Zagreb,
 Ka\v{c}i\'ceva 26, HR-10000 Zagreb, Croatia}
\author{B. R. Dennis}
\affil{NASA Goddard Space Flight Center, Greenbelt, MD 20771, U.S.A.}
\author[1]{M. Temmer}
\author[1]{A. Hanslmeier}
\author[2,4]{J. Magdaleni\'{c}}
\affil{Trieste Astronomical Observatory, Via G.B.\ Tiepolo 11, I-34131 Trieste, Italy}

\begin{document}

%\keywords{\LaTeX; ESA; macros}

\maketitle

\begin{abstract}
Based on simultaneous observations of solar flares in hard
and soft X-rays we studied several aspects of the Neupert effect.
About half of 1114 analyzed  events show a timing behavior
consistent with the \mbox{Neupert} effect. For these events,
a high correlation between the soft \mbox{X-ray} peak flux and
the hard \mbox{X-ray} fluence is obtained, being indicative of
electron-beam-driven evaporation. However, for about one fourth
of the events there is strong evidence for an additional heating
agent other than electron beams. We discuss the relevance of these
findings with respect to Parker's idea of coronal heating by nanoflares.
\end{abstract}

\section{Introduction}

The Neupert effect is the name given to the observational finding that
the rising part of the soft \mbox{X-ray} (SXR) light curve often resembles the
time integral of the hard X-ray (HXR) or microwave emission (\mbox{Neupert},
1968; Dennis \& Zarro, 1993). The physical relevance of the Neupert effect basically
arises from the fact that it is interpreted as a causal connection between
the thermal and nonthermal flare emissions, which can be naturally explained
within the nonthermal thick-target model (Brown, 1971).
In this model, the flare energy is released primarily in the form
of nonthermal electrons, and hard X-rays are produced via electron-ion
bremsstrahlung when the electron beams impinge on the
lower corona, transition region and chromosphere. The
model assumes that only a small fraction of the electron beam
energy is lost through radiation; most of the loss is due to
Coulomb collisions that serve to heat the ambient plasma.
As a consequence of the rapid energy deposition
a strong pressure imbalance develops between the dense, heated chromosphere
and the tenuous corona. The high pressure gradients cause the
heated plasma to convect into the corona in a process known as
chromospheric evaporation (Antonucci et al.\ 1984; Fisher et al.\ 1985),
where it gives rise to enhanced SXR emission via thermal
bremsstrahlung.

In this case, the hard X-ray flux is linked to the instantaneous rate of
energy supplied by electron beams, whereas the soft X-ray flux is
related to the accumulated energy deposited by the same electrons up to
that time, and we can expect to see the Neupert effect. Any deviation from
the Neupert effect, in principle, suggests that the hot SXR emitting plasma
is not heated exclusively by thermalization of the accelerated electrons
that are responsible for the HXR emission. Therefore, investigations of
the Neupert effect provide insight into the role of nonthermal
electrons for the flare energetics.

The Neupert effect can be expressed as %(e.g., Lee et al. 1995)
\begin{equation}
F_{\rm P,SXR} = k \cdot {\cal F}_{\rm HXR}  , \label{EqNeup}
\end{equation}
with $F_{\rm P,SXR}$ the SXR peak flux and ${\cal F}_{\rm HXR}$
the HXR fluence, i.e.\ the HXR flux integrated over the event duration.
The coefficient~$k$ depends on several factors, as, e.g., the magnetic
field geometry and the viewing angle, and thus may vary from flare to flare
(Lee et al., 1995). However, if $k$ does not depend systematically on the
flare intensity, then the SXR peak flux and the HXR fluence are linearly related.
% linear relationship is expected to exist between the

\section{Data Selection}

We utilize the SXR data from the {\it Geostationary Operational
Environmental Satellites} (GOES) and the HXR data from the
{\it Burst and Transient Source Experiment} (BATSE) aboard the
{\it Compton Gamma Ray Observatory}. The X-ray sensor aboard GOES
consists of two ion chamber detectors, which provide whole-sun X-ray
fluxes in the \mbox{0.5--4} and 1--8~{\AA} wavelength bands.
BATSE is a whole-sky HXR flux monitor that, in part, consists of
eight large-area detectors. From each detector there are hard X-ray
measurements in four energy channels, 25--50, \mbox{50--100}, 100--300
and $>$300~keV (Schwartz et al., 1992).
%A characteristics of the BATSE instrument and its
%capabilities for solar flare studies can be found in Fishman
%et al.\ (1992) and Schwartz et al.\ (1992).

For the analysis, the 1-min averaged GOES SXR data measured in the
1--8~{\AA} channel and the HXR data collected in the BATSE Solar Flare
Catalog, archived in the Solar Data Analysis Center at NASA/Goddard
Space Flight Center for the period 01/1997\,--\,06/2000 are used.
The peak and total count rates are
background subtracted for the flux below 100~keV.
For the SXR events, we used the flux just before the
flare start for background \nolinebreak \mbox{subtraction}.
To be identified as corresponding events we demand that
the start time difference between a SXR and a HXR event does not
exceed 10~min. Overlapping events are excluded. Applying these
criteria, we obtained 1114 events that were observed in
both hard and soft X-rays (for details see Veronig et al.,
2002a).

\begin{figure}[tb]
\centering
\hspace*{-0.2cm}
\resizebox{0.94\hsize}{11.cm}{\includegraphics{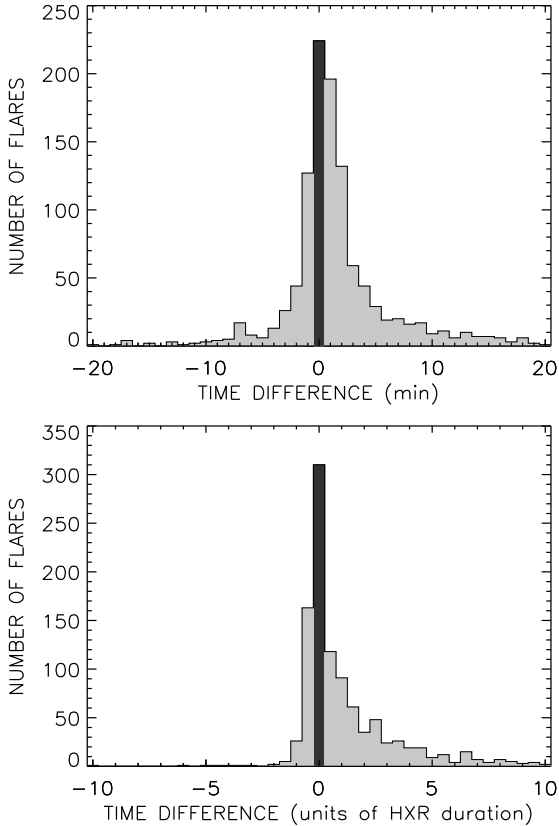}}
    \caption[]{Histogram of the difference of the SXR maximum and
    the HXR end time in absolute values (top panel) and
    normalized to the HXR event duration (bottom panel). Positive
    values indicate that the maximum of the SXR emission occurs
    after the end of the HXR emission, negative values vice versa.
    \label{Fig_timing}}
\end{figure}

\section{Analysis}

For each event we determined the difference, $\Delta t$, of the peak time of the SXR
emission, $t_{\rm SXR,P}$, and the end time of the HXR emission, $t_{\rm HXR,E}$.
Furthermore, the time differences were normalized to the
duration~$D$ of the respective HXR event, i.e.\
\begin{equation}
\Delta t_{\rm norm} = \frac{\Delta t}{D} = \frac{t_{\rm SXR,P} - t_{\rm HXR,E}}{D} \, .
\end{equation}
Figure~\ref{Fig_timing} shows the histogram of the absolute and
normalized time differences. Both representations of the SXR--HXR time
difference have its mode at zero.
49\% of the events lie within the range $|\Delta t| \leq 1$~min,
and 65\% within $|\Delta t| \leq 2$~min. For the normalized time differences,
we obtain that 44\% lie within $|\Delta t_{\rm norm}| \leq 0.5$~HXR units, and
59\% within $|\Delta t_{\rm norm}| \leq 1$~HXR unit. This outcome suggests that
certainly a considerable part of the events coincides well with the expectations
from the \mbox{Neupert} effect regarding the relative timing. %of the SXR and HXR emission.

\begin{figure}
\centering
\resizebox{0.98\hsize}{!}{\includegraphics{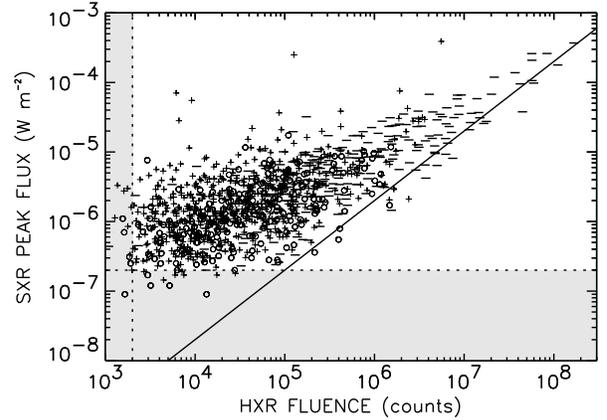}}
    \caption[]{Scatter plot of the SXR peak flux versus the HXR
    fluence. % for the complete sample of 1114~events.
    The regions that lie beyond the estimated HXR fluence and SXR peak flux
    thresholds are grey shaded. Moreover, we have indicated
    the sign of the time difference between SXR peak and HXR end
    for each single event: {\rm ``$+$"} symbols denote events with positive,
    {\rm ``$-$"} symbols events with negative, {\rm ``$\circ$"} symbols events
    with zero time difference (i.e.\ the SXR peak and HXR end take place within $1$\,min).
    The straight line indicates a line of constant~$k$,
    i.e.\ $F_{\rm P,SXR} = 2\cdot 10^{-12} \cdot {\cal F}_{\rm HXR}$.
    \label{Fig_FluencePeak} }
\end{figure}

Figure~\ref{Fig_FluencePeak} shows the scatter plot of the SXR
peak flux versus the HXR fluence for the complete sample, clearly
revealing an increase of $F_{\rm P,SXR}$ with increasing ${\cal
F}_{\rm HXR}$. It can also be inferred from the figure that the
slope is not constant over the whole range but that it is larger
for large HXR fluences than for small ones. For very large fluences,
the slope approaches the value of~1, indicative of a linear relation
between the SXR peak flux and HXR fluence.
We stress that the slope at small fluences might be affected by missing
events with small SXR peak fluxes (due to selection effects),
and thus appear flatter than it is in fact.
Figure~\ref{Fig_FluencePeak} reveals an interdependence between the importance
of an event and the sign of the time difference. Basically all large
flares belong to the group of events with $\Delta t < 0$, i.e.\ the
SXR peak occurs before the HXR end. Moreover, the flares with $\Delta t < 0$
reveal a strong tendency to be of long duration.

We obtain a high cross-correlation coefficient for the
SXR peak flux and HXR fluence relationship, $r=0.71$. This coefficient is
higher than those for the SXR peak flux and HXR peak flux,
$r=0.57$. This indicates that the correlation is primarily due to the
HXR fluence -- SXR peak flux relationship, as predicted from the
Neupert effect, and not, e.g., due to the fact that flares
with large HXR peak fluxes also tend to have intense SXR counterparts.
Furthermore, it is important to note that the HXR fluence~--
SXR peak flux correlation is higher for the
events with negative time differences, $r = 0.82$, than for the
events with positive time differences, $r = 0.54$.

On the basis of the relative timing of the SXR peak and the HXR
end, we extracted two subsets of events. The events of set~1 are roughly
consistent with the timing expectations of the Neupert effect,
and the events of set~2 are inconsistent with it.
The two sets are defined by the following conditions:
\begin{eqnarray*}
{\rm Set~1:~} (|\Delta t| < 1 {\rm ~min}) & {\rm \hspace*{-0.2cm} OR}   & \hspace*{-0.2cm} (|\Delta
t_{\rm norm}| < 0.5{\rm ~unit}) \, ,\\
{\rm Set~2:~} (|\Delta t| > 2 {\rm ~min}) & \hspace*{-0.2cm} {\rm AND}  & \hspace*{-0.2cm} (|\Delta
t_{\rm norm}| > 1.0 {\rm~unit})\, .
\end{eqnarray*}
The applied conditions represent a combination of absolute and
normalized time differences in order to avoid as much as possible
any a priori interdependence with the flare duration and/or flare
intensity. Out of the 1114 corresponding HXR and SXR flares, 485 (44\%)
events fulfilled the timing criterion of set~1; 270~events (24\%)
belong to set~2; 359 events (32\%) are neither attributed to set~1 nor
to set~2.

\begin{figure}[tb]
\centering
\resizebox{0.98\hsize}{11.1cm}{\includegraphics{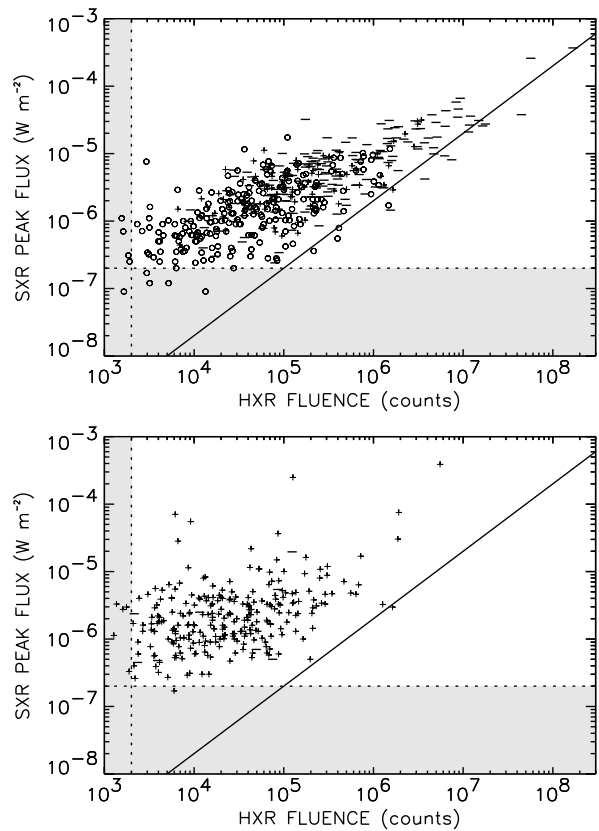}}
    \caption[]{Scatter plot of the SXR peak flux versus the HXR fluence
    separately plotted for set~1 (top panel) and set~2 (bottom panel).
    The straight lines are lines of constant~$k$ $(=2\cdot 10^{-12})$.
    \label{Fig_FluencePeak2} }
\end{figure}

In Figure~\ref{Fig_FluencePeak2}, we plot the SXR peak flux versus
the HXR fluence separately for set~1 and set~2. The figure reveals
that the two sets have very different characteristics besides the different
timing behavior. Set~1 contains many more large events and shows a steeper
increase of $F_{\rm P,SXR}$ with increasing ${\cal F}_{\rm HXR}$ than set~2.
Moreover, set~1 contains more events with negative than
positive time difference, whereas almost all events of set~2
are characterized by $\Delta t > 0$, i.e.\ increasing SXR emission
beyond the end of the hard X-rays. On average, for small fluences the events
belonging to set~2 have a larger SXR peak flux at a given HXR fluence
than those of set~1, indicating an ``excess" of SXR emission with
respect to set~1. The cross-correlation coefficients derived separately
for the subsets reveal that the correlation among the SXR peak flux and
HXR fluence is much more pronounced for the events of set~1, $r=0.78$,
than those of set~2, $r=0.41$.

\begin{figure}[tb]
\centering
\hspace*{0.03cm}
\resizebox{0.94\hsize}{10.3cm}{\includegraphics{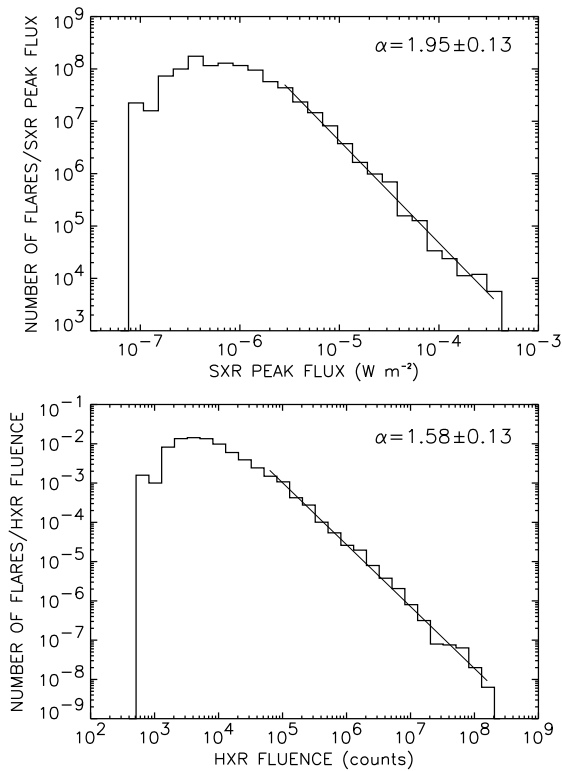}}
    \caption[]{Flare frequency distributions as function of the SXR peak
    flux (top panel) and HXR fluence (bottom panel). The straight line
    indicates the linear fit in log-log space, characterized by a slope $-\alpha$.}
    \label{Fig_distr}
\end{figure}

\section{Discussion and Conclusions}

24\% of the events have $\Delta t < 0$, i.e.\ the SXR
maximum occurs before the end of the HXR emission.
These events are preferentially of long duration.
Li et al.\ (1993) have calculated time profiles of soft
and hard X-ray emission from a thick-target
electron-heated model, finding that, in general, the time derivative
of the SXR emission corresponds to the time profile of the HXR
emission, as stated by the Neupert effect. However, for gradual
events they obtained that this relationship breaks down during
the decay phase of the HXR event, in that the maximum of the SXR
emission occurs before the end of the HXR emission. This phenomenon
can be explained by the fact that the SXR emission starts
to decrease if the evaporation-driven energy supply cannot overcome the
instantaneous cooling of the hot plasma, which is likely to
happen in gradual flares. Considering our observational findings
together with the results from simulations by Li et al.\ (1993),
presumably most of the events with $\Delta t < 0$ are consistent with the
electron-beam-driven evaporation model. In particular, the high
correlation between $F_{\rm P,SXR}$ and ${\cal F}_{\rm HXR}$,
$r \approx 0.8$, supports such interpretation.

56\% of the events have $\Delta t > 0$; these events are preferentially
of short and weak HXR emission. In principle, the fact that the SXR
emission is still increasing although the HXR emission, i.e.\ the
electron input, has already stopped indicates that an additional agent
besides the HXR emitting electrons is contributing to the energy input and
prolonging the heating and/or evaporation. However, McTiernan et
al.\ (1999) have shown that the SXR time profile depends on the
temperature response of the used detector:
An increase of the SXR emission of low-temperature flare plasma after the HXR
end may arise due to cooling of high-temperature plasma. Thus, we cannot
attribute all events with $\Delta t > 0$ as inconsistent with the
electron-beam-driven chromospheric evaporation model.
Instead, we consider as inconsistent only flares which show strong deviations
from $\Delta t = 0$, i.e.\ the events belonging to set~2.

This means that for at least one fourth of the analyzed events an
additional heating agent besides nonthermal electrons is
suggested. A probable scenario is that energy is transported
from the primary energy release site via thermal conduction fronts,
which initiate chromospheric evaporation but do not produce hard \mbox{X-rays}.
The finding that for a considerable fraction of flares, preferentially
weak ones, an additional heating agent other than electron beams is
suggested, is not only relevant for the flare energetics but also
for Parker's idea of coronal heating by nanoflares (Parker, 1988).

Hudson (1991) pointed out that if the corona is heated by flare-like events
of different sizes, then the flare energy distribution must have a
power-law slope $\alpha > 2$. If the SXR flux does not vary systematically
with temperature and density, then the SXR peak flux is
linearly related to the maximum thermal energy of the flare plasma
(see Lee et al., 1995; Veronig et al., 2002a). On the other hand, HXR fluence
distributions can be considered as representative for the energy contained
in nonthermal electrons. In Figure~\ref{Fig_distr}, we show the SXR
peak flux and the HXR fluence distributions derived from 1114
corresponding SXR/HXR flares, finding $\alpha = 1.95$\,$\pm$\,0.13
and $\alpha = 1.58$\,$\pm$\,0.13, respectively. The discrepancy
between the slopes of the HXR fluence and the SXR peak flux distributions
was already pointed out and discussed in Lee et al.\ (1993, 1995) and
Veronig et al.\ (2002b). The present analysis
provides an explanation for this difference in power-law slopes:
The relationship between the SXR peak flux and the HXR fluence
is not linear, %conflicting with the electron-heated thick-target model,
whereby the deviations from a linear correlation are strongest for weak
flares (cf.~Figure~\ref{Fig_FluencePeak}).

The soft X-ray flare emission increases due to energy supply by electron
beams as well as any other heating agent, whereas the hard X-ray
emission contains only information on the energy provided by
electrons. Together with our finding that particularly in weak events
an additional heating agent besides electron beams is suggested,
this strongly suggests that soft X-ray peak flux distributions are a
more meaningful indicator of flare energy distributions than hard X-ray
fluence distributions. Furthermore, we have shown that weak flares have
different characteristics than large flares, in the sense that electrons
are less important for their energetics.
%Furthermore, to infer on the contribution of nanoflares to the
%heating of the corona, the power-law index of the frequency distribution
%of observed flares is extrapolated to flare sizes that lie below
%the observational limits, i.e.\ nanoflares. However, this extrapolation
%is only valid under the presumption that nanoflares have the same
%characteristics than the observed, more intense, flares.
%We have shown that weak flares have different characteristics than
%large flares, in that sense that nonthermal electron beams are
%less important for their energetics.
%
Therefore, it is possible that the power-law slope of
nanoflare frequency distributions differs from that derived for
observed flares, which is close to the critical value of~$2$.
In this respect it should be worthwhile to investigate flare frequency
distributions of SXR flares without HXR counterparts, i.e.\ without
detectable particle acceleration, since these flares possibly provide the
link to the smallest flare-like energy release events. %reconnection events.

\section*{Acknowledgements}
A.~V., M.~T.\ and A.~H.\ gratefully \mbox{acknowledge} the Austrian
{\em Fonds zur F\"orderung der wissen\-schaft\-lichen
Forschung} (FWF grant P15344-PHY) for supporting this project.

\end{document}